\documentclass[aps,article,prc,preprintnumbers,twocolumn]{revtex4}

\usepackage{epsfig}
\usepackage{dcolumn}
\usepackage{bm}
\usepackage{multirow}

\begin{document}
\title{Examining a reduced jet-medium coupling in Pb+Pb collisions at the Large Hadron Collider}

\author{Barbara Betz$^{a}$}
\author{Miklos Gyulassy$^b$}
\affiliation{$^a$Institute for Theoretical Physics,
Johann Wolfgang Goethe-University, 60438 Frankfurt am Main,
Germany\\
$^b$Department of Physics, Columbia University, 
New York, 10027, USA}

\begin{abstract}

Recent data on the nuclear modification factor $R_{AA}$ of jet fragments 
in $2.76$~ATeV Pb+Pb collisions at the Large Hadron Collider (LHC) 
indicate that the jet-medium coupling in a Quark-Gluon Plasma (QGP) is 
reduced at LHC energies and not compatible with the coupling deduced
from data at the Relativistic Hadron Collider (RHIC). We estimate the 
reduction factor from a combined fit to the available data on 
$R_{AA}(\sqrt{s},p_T,b)$ and the elliptic flow $v_2(\sqrt{s},p_T,b)$ at 
$\sqrt{s}=0.2,2.76$ ATeV over a transverse momentum range $p_T\sim 10-100$~GeV 
and a broad impact parameter, $b$, range. We use a simple analytic ``polytrope'' 
model ($dE/dx=- \kappa E^{a} x^z T^{c}$) to investigate the dynamical jet-energy 
loss model dependence. Varying $a=0-1$ interpolates between weakly-coupled and 
strongly-coupled models of jet-energy dependence while $z=0-2$ covers a wide 
range of possible jet-path dependencies from elastic and radiative to 
holographic string mechanisms. Our fit to LHC data indicates an approximate 40\% 
reduction of the coupling $\kappa$ from RHIC to LHC and excludes
energy-loss models characterized by a jet-energy exponent with $a>1/3$.
In particular, the rapid rise of $R_{AA}$ with $p_T\ge 10$~GeV 
combined with the slow variation of the asymptotic $v_2(p_T)$ at the LHC 
rules out popular exponential geometric optics models ($a=1$). The LHC data 
are compatible with $0\leq a\leq 1/3$ pQCD-like energy-loss models where the
jet-medium coupling is reduced by approximately 10\% between RHIC and LHC.

\end{abstract}

\maketitle

\section{Introduction}

First data from the Large Hadron Collider (LHC) on the nuclear-size 
dependence of jet-medium interactions in Pb+Pb collisions at $\sqrt{s}=2.76$ ATeV 
\cite{Aamodt:2010jd,Otwinowski:2011gq,Appelshauser:2011ds,collaboration:2011zi,Schukraft:2011cz,Dainese:2011vb,ATLAS:2011ag,Steinberg:2011dj,collaboration:2011hfa,CMSPas,collaboration:2011qp,Chatrchyan:2011pb,Chatrchyan:2011eka,Chatrchyan:2011sx}
showed  that the jet-medium coupling at the LHC is weaker than expected 
\cite{Horowitz:2011gd,Zakharov:2011ws,Buzzatti:2011vt,Vitev:2011gs} 
from fixed coupling extrapolations from $\sqrt{s}=0.2$ ATeV data
at the Relativistic Heavy Ion Collider (RHIC)
\cite{sqgp,whitebrahms,whitephobos,whitestar,whitephenix}.
From the factor of $\sim 2$ increase of global multiplicity per unit 
rapidity, $dN_{\rm ch}/d\eta\approx 1600$ in central Pb+Pb reactions at 
the LHC relative to RHIC, substantially more suppression of high-$p_T$ 
pions was predicted than observed. In this paper we estimate the reduction of 
the jet-medium coupling implied by the new data and test the consistency
for a wide variety of jet-energy loss models. 

An open question from studies of the nuclear modification factor of jets at 
RHIC is whether jet-medium interactions in dense deconfined Quark-Gluon 
Plasma (QGP)-matter can be better described in terms of weakly-coupled 
perturbative QCD (pQCD) tomography or novel strongly-coupled gravity-dual 
Anti-de-Sitter/Conformal Field Theory (AdS/CFT) string-model holography 
\cite{Maldacena:1997re,coester,MGtomoholo}. The first data from LHC 
provide stringent new tests of jet-medium interaction models for a 
higher QGP-density range and for an order of magnitude higher transverse 
momentum range. The doubling of the QGP density at LHC relative to RHIC 
extends the temperature range by approximately 30\% beyond the range 
explored at RHIC.

In order to extract either tomographic or holographic information
from the jet-quenching observable systematics, it is important to specify
the initial jet flux at the jet-production points and the initial geometry 
of the QGP medium. Two competing models, the Glauber \cite{glauber} model
and the Color Glass Condensate (CGC) \cite{cgc}, have been used so far 
describe the initial QGP medium geometry and to predict the impact 
parameter $b$ and beam energy $\sqrt{s}$ dependence of those distributions. 
The Glauber model is based on an eikonal Wood-Saxon nuclear geometry and 
assumes incoherent superposition of proton-proton collisions while CGC models
are based on non-linear, small-x saturation effects \cite{kln,adrian,dumitru}. 
Here, we calculate jet-quenching observables using both initial geometries
as a measure of current systematic theoretical errors associated with uncertainties 
of the initial QGP geometries as a function of impact parameter.

In this paper, we use a simplified analytic ``polytrope'' jet-energy-loss model 
\cite{Betz:2011tu,Betz:2011jp} that can interpolate between a wide 
class of weakly and strongly-coupled models of jet-medium interaction in high-energy 
nuclear collisions:
\begin{eqnarray}
\hspace*{-3ex}
\frac{dP}{d\tau}(\vec{x}_0,\phi,\tau)= 
-\kappa P^a(\tau) \, \tau^{z} \, T^c[\vec{x}_\perp(\tau),\tau ;b]\;.
\label{GenericEloss}
\end{eqnarray}
The energy loss per unit length is characterized by the three exponents ($a,z,c$)
that determine the jet-energy dependence $P^a$, the path-length dependence 
$\tau^z$, and the local temperature-power dependence $T^c(\vec{x}_\perp,\tau)$.
This polytrope model reproduces the results of detailed Djordjevic-Gyulassy-Levai-Vitev 
(DGLV) opacity series calculations remarkably well \cite{Horowitz:2011gd,Buzzatti:2011vt} 
but has the advantage of allowing quick estimates of the variations of predictions 
to even large deformations of dynamical assumptions.

\begin{figure*}
\hspace*{-0.8cm}
\includegraphics[scale=0.5]{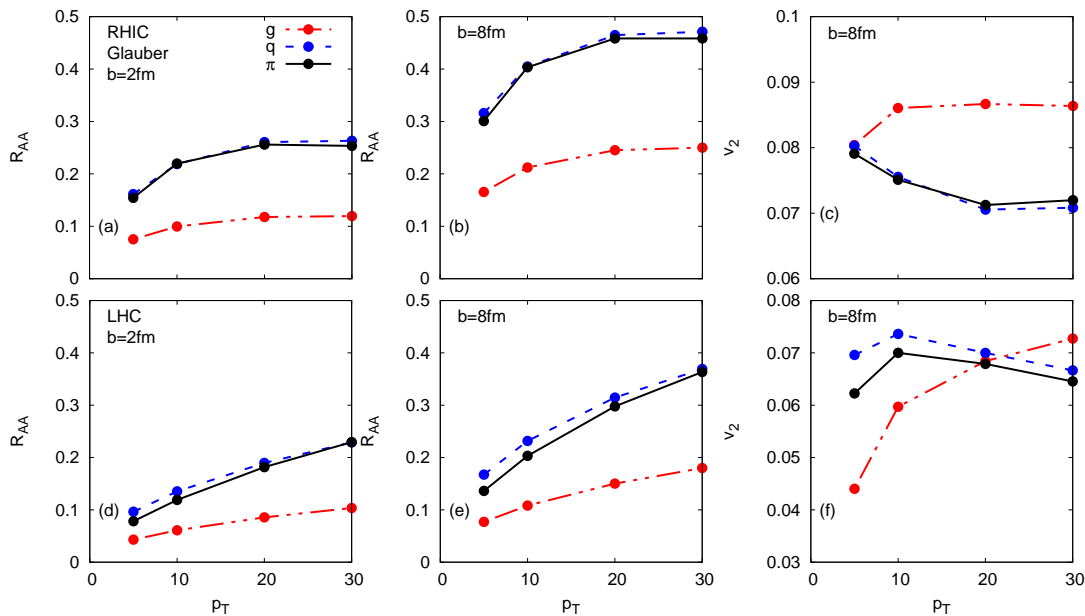}
\caption{(Color online) Illustration of the remarkable similarity
of quenched pion $R_{AA}$ and $v_2$ (solid black line) to the underlying
quark-jet nuclear modification (blue dashed lines) and the approximate 
independence of the pion nuclear modification factor on the much higher quenched 
gluon jets (red dashed-dotted lines) at both RHIC (upper panels) and LHC
(lower panels). Left panels correspond to central $b=2$~fm while 
center and right panels refer to $b=8$~fm. A polytrope energy loss with 
($a=1/3,z=1,c=8/3$) is assumed.}
\label{fig1}
\end{figure*}

The energy loss per unit length, $dE/dx=dP/d\tau$ depends on the local proper time 
$\tau$ in a frame where the jet rapidity $y$ vanishes. The jet path is thus 
perpendicular to the beam axis in that frame and can be assumed to be a straight eikonal line
$\vec{x}_\perp(\tau)=\vec{x}_0+\hat{n}(\phi)\tau$ from the production point 
$\vec{x}_0$ in direction of the jet azimuthal angle $\phi$ relative to the reaction 
plane $\hat{b}$. Bjorken longitudinal expansion \cite{bjorken} is taken into account 
by $T(\vec{x},\tau)= T(\vec{x},\tau_0)(\tau_0/\tau)^{1/3}$ until a freeze-out 
isotherm is reached specified by $T[\vec{x}(\tau_f),\tau_f;b]=T_f=100$~MeV. As in Refs.\ 
\cite{Betz:2011tu,Betz:2011jp}, we assume that the energy loss depends monotonically 
on the co-moving local entropy density. We consider an initial QGP-formation time of 
$\tau_0=1$ fm/c \cite{heinzrecent}. 

The dimensionless effective jet-medium coupling $\kappa$ is interpreted as proportional 
to $\alpha_s^3$ in the case of radiative energy-loss tomography while it is interpreted to be 
proportional to $\kappa \propto \sqrt{\lambda_{tH}}\propto (\alpha_s N_c)^{1/2}$ in terms of 
the t'Hooft coupling in gravity dual string holography. 

We note that the considered monotonic power-law dependence of $dE/dx$ on the temperature 
(or entropy) field in Eq.\ (\ref{GenericEloss}) is a dynamical assumption consistent with
pQCD tomography as well as AdS/CFT holography models in the literature. However, there 
also exist non-monotonic models, e.g. see Liao and Shuryak \cite{Liao:2011kr} involving 
e.g.\ magnetic monopole enhanced $dE/dx$ near the critical QCD transition point of 
$T_c\approx 170$ MeV. The scope of the present paper is limited to monotonic models as 
in Eq.\ (\ref{GenericEloss}) to avoid extra model assumptions and parameters.

We also note that the $(a=1/3, z=1, c=8/3)$ polytrope model describes approximately both the 
pQCD and the AdS/CFT falling string cases \cite{ches1,ches2}. An $(E/T)^{1/3}$-energy 
dependence is numerically similar to the logarithmic $\log(E/T)$ dependence predicted 
by fixed coupling pQCD-energy loss in the range $10<E/T<600$ relevant at LHC energies. 
This power law is also predicted to be the lower bound of the power $a$ in the 
falling-string scenario in an AdS/CFT conformal holography.

Variants of the polytrope model with path dependencies varying from $z=1$ to $z=2$ have 
been considered in Refs.\ \cite{Jia,Adare:2010sp,Fries:2010jd,ches1,ches2,ches3,Arnold:2011qi,Ficnar:2011yj}.
Recent work \cite{Ficnarnew} on the falling string energy-loss in conformal AdS 
geometry has identified important corrections to the original works \cite{ches1,ches2} 
that may reduce the effective $z=2$ path-length power-law dependence assumed in 
\cite{Jia} to $z\approx 1$ similar to the predicted radiative energy loss in pQCD. 
While such a falling string scenario essentially leads to the same polytrope exponents as radiative  
pQCD-energy loss, the jet-medium coupling strength differs significantly in the two cases. 

As we also show below, Eq.\ (\ref{GenericEloss}) reduces to the survival probability 
model referred to as "AdS/CFT"-like in Refs.\ \cite{Jia,Adare:2010sp,horojia} in the 
limit $(a=1,z=2,c=3)$. For $(a=1, z=0,c=2)$, on the other hand, the polytrope model 
reduces to the heavy-quark string-drag energy-loss of conformal AdS holography \cite{ches1,ches2}. 
As we emphasize below, polytropes with $a=1$ lead in fact to a generalized class of 
``geometric optics'' or ``survival probability'' models of nuclear jet modifications. 
We will demonstrate that the recent LHC data on $R_{AA}(p_T)$ data rules out this 
class of dynamical models.
 
In this work, we further generalize Ref.\ \cite{Betz:2011tu} by using 
the full pQCD jet-production $p_T$-spectra for the initial quark and gluon jets 
avoiding the local spectral index approximation. In addition, we convolute the 
quenched jet spectra with KKP pion fragmentation functions \cite{Kniehl:2000hk} 
that have been tested well against RHIC $pp\rightarrow \pi^0$ spectra in Ref.\ \cite{Simon:2006xt}. 

The magnitude of the cube of the initial temperature profile is assumed to scale with 
the observed rapidity density $T_0^3(\vec{x},b) \propto \rho(\vec{x},b) dN_{ch}(\sqrt{s},b)/d\eta$. 
The initial transverse coordinate distribution, $\rho(\vec{x},b)$, of the QGP is modelled 
according to a Glauber and a higher eccentricity CGC-like elliptically deformed geometry
[see Eqs.(\ref{fkln1}) to (\ref{fkln2}) below].

\begin{figure}
\hspace*{-1.1cm}
\vspace*{-0.3cm}
\includegraphics[scale=0.535]{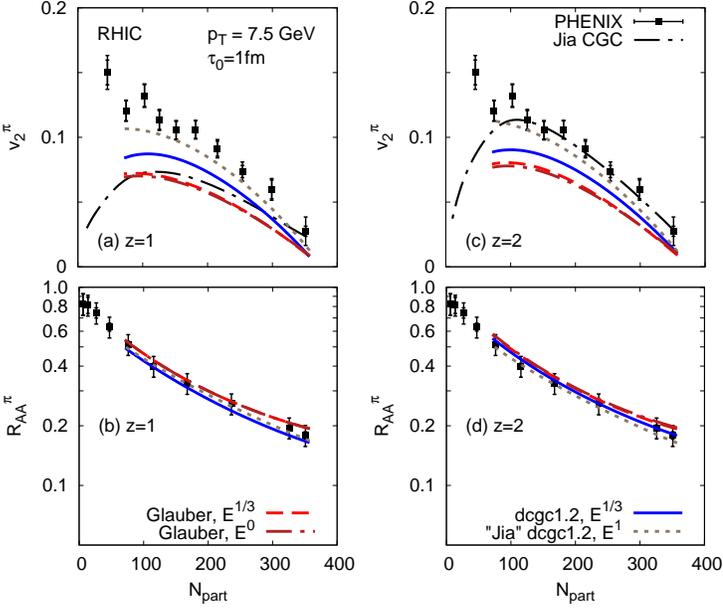}
\caption{(Color online) The $R^\pi_{AA}(N_{part})$ and the 
$v^\pi_2(N_{part})$ at RHIC energies after fragmentation for
an energy loss with $z=1$ (left panel) as well as an energy loss 
with $z=2$ (right panel). Glauber initial conditions are displayed by 
the red dashed lines and dcgc1.2 initial conditions by the blue solid 
lines. Those results are obtained for $a=1/3$. The grey dotted 
line represents an energy-loss for dcgc1.2 initial 
conditions considering just binary collisions (``Jia'' dcgc1.2) as well as $a=1$, 
similar to the one in Refs.\ \cite{Jia,Adare:2010sp}.
The dark red dashed-dotted line displays an energy loss with 
Glauber initial conditions and $a=0$. All calculations assume an 
initialization time of $\tau_0=1$~fm. The black dashed-dotted line 
is the result for CGC initial conditions from Ref.\
\cite{Adare:2010sp} with $\tau_0=0$~fm. Data are taken from Ref.\ 
\cite{Adare:2010sp}.}
\label{fig2}
\end{figure}

\section{A Polytrope Model of Jet-Quenching}
At the partonic level, the nuclear modification factor $R_{AA}$ is the 
ratio of the jet spectrum for jets penetrating a QGP produced in A+A collisions 
to the initial jet spectrum predicted by pQCD without final state
interactions:
\begin{eqnarray}
R_{AA}^{q,g}(P_f,\vec{x}_0,\phi)&=&\frac{dN^{jet}_{QGP}(P_f)}{dyd\phi dP^2_f}{\Big/ }
\frac{dN^{jet}_{vac}(P_f)}{dyd\phi dP^2_0}\nonumber\\
&&\hspace*{-1.4cm}=\frac{dP^2_0}{dP^2_f}
\frac{dN^{jet}_{vac}[P_0(P_f)]}{dyd\phi dP_0^2}{\Big/ }
\frac{dN^{jet}_{vac}(P_f)}{dyd\phi dP_0^2}\,.
\end{eqnarray}
Denoting the invariant jet distribution by $g_0(P)$,
\begin{eqnarray}
g_0(P)&=&\frac{dN^{jet}_{vac}(P)}{dyd\phi dP^2}\,,
\end{eqnarray}
the nuclear modification factor for a quark (q) or gluon (g) jet 
with a final momentum $P_f$, produced at a transverse coordinate $\vec{x}_0$ 
and propagating in direction $\phi$ is, from Eq.\ (\ref{GenericEloss}),
\begin{eqnarray}
R_{AA}^{q,g}(P_f,\vec{x}_0,\phi)&=&\frac{g_0[P_0(P_f)]}{g_0(P_f)}
\frac{dP^2_0}{dP^2_f}\,.
\label{RAAqg}
\end{eqnarray}

\begin{figure}
\hspace*{-0.6cm}
\vspace*{-0.3cm}
\includegraphics[scale=0.5]{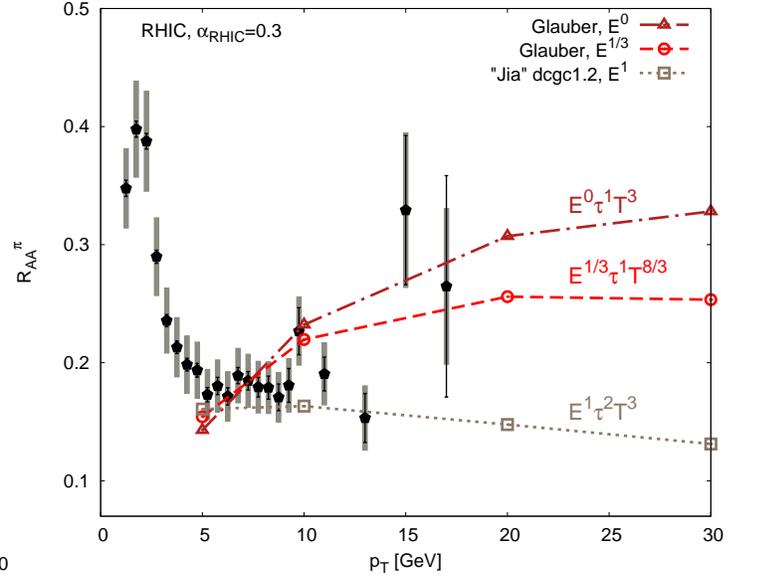}
\caption{(Color online) The $R^\pi_{AA}$ at $0-5\%$ centrality as a function of  
$p_T$ at RHIC energies. The grey dotted line shows an energy loss for dcgc1.2 initial 
conditions considering just binary collisions (``Jia'' dcgc1.2) and ($a=1,z=2,c=3$), while the red dashed and dark red
dashed-dotted lines dispay Glauber initial conditions for ($a=1/3,z=1,c=8/3$) and 
($a=0,z=1,c=3$), respectively. The initialization time is chosen to be $\tau_0=1$~fm. 
The data are taken from Ref. \cite{PHENIX}.}
\label{fig3}
\end{figure}

The polytrope model introduced in Eq.\ (\ref{GenericEloss}) is convenient because
the initial jet-parton momentum, $P_0(P_f)$ depends on the final quenched 
parton momentum $P_f$ analytically as \cite{Horowitz:2011gd,Betz:2011tu}
\begin{eqnarray} 
\hspace*{-1ex} 
P_0(P_f)=\left[P_f^{1-a}+{K}\int_{\tau_0}^{\tau_f}
\tau^z T^{c}[\vec{x}_\perp(\tau),\tau]d\tau
\right]^\frac{1}{1-a}\hspace*{-2ex},
\label{FinMom}
\end{eqnarray}
where $K=(1-a) \kappa C_2$ for gluon(quark) jets. Non-monotonic 
density-dependent scenarios as in Ref.\ \cite{Liao:2011kr}
can be simulated by introducing an additional local temperature-dependent
function $f(T)$ inside the path integral. However, as noted in the introduction, 
we limit our applications to monotonic temperature dependencies of the energy loss 
per unit length given by $T^c$.

Note that if $\kappa$ is dimensionless and no additional dimensionful scales 
influence the energy loss, the temperature exponent depends on $a,c$. 
Any deviations from the dimensional constraint
\begin{equation}
c=2-a+z
\label{sumrule}
\end{equation}
found by fitting jet systematics can be used to help identify the existence of other 
relevant dimensionful scales. Non-conformal physics that may modify energy loss 
near the cross-over temperature $T_c$ or running coupling effects depending on 
$\Lambda_{QCD}\sim T_c$ could lead to violations of Eq.(\ref{sumrule}). 
In case of heavy-quark drag holography, ($a=1,z=0,c=2$), the sum rule is violated 
because of occurrence of an extra $1/M_Q$ factor of dimension $-1$ \cite{ches1,ches2}.

The limit with $a=1$ is of special interest and used frequently in the literature 
because it leads to a pure exponential dependence of $P_0(P_f)$:
\begin{equation}
P_0(P_f)=P_f \; e^{\chi_{z,c}}  
\label{a1p0} 
\end{equation}
with a {\em jet-energy independent} effective opacity 
\begin{equation}
\chi_{z,c}(\phi)= \kappa C_2\int_{\tau_0}^{\tau_f} d\tau \tau^z T^c(\tau,\phi)
\,.\label{chizc} 
\end{equation}
This corresponds to a generalized ``geometric optics'' limit
with $c=1+z$ if no other relevant scales are involved.

In this particular case, the nuclear modification factor at the parton level 
then reduces to
\begin{equation}
R_{AA}= g_0(P_f \; e^\chi)\; e^{2 \chi}/g_0(P_f)\,.
\label{a1lim}
\end{equation}
Approximating the initial jet spectrum by a simple
power law, $g_0\propto p^{-n}$, where $n$ is the jet spectral index,
the nuclear modification factor reduces to the simplest possible
{\em energy independent} ``jet survival probability'' in azimuthal 
direction $\phi$ given by
\begin{eqnarray}
R_{AA}(\phi)= e^{(2-n)\chi_{z,c}(\phi)}=e^{-\chi_{eff}(\phi)} \equiv P_0(\phi)\,.
\label{a1limlim}
\end{eqnarray}

Since at both RHIC and LHC the spectral indices of quarks and gluons differ and 
vary significantly with the jet energy, we must use Eq.\ (\ref{a1lim}) rather than 
the naive constraint of Eq.\ (\ref{a1limlim}) to explore the consequences of 
$a=1$ energy loss models.

For the general $(a,z,c)$ case, we use Eqs.\ (\ref{RAAqg}) and (\ref{FinMom}) to compute
the parton level nuclear modification factor. The final pion nuclear modification reads
\vspace*{1ex}
\begin{widetext}
\begin{eqnarray}
\hspace*{0.8cm}
R_{AA}^\pi(p_\pi,\phi) &=& 
=\frac{\sum\limits_{\alpha=q,g}\;\int\limits_{z_{min}}^1 \frac{dz}{z} \; d\sigma_\alpha\left(\frac{p_\pi}{z}\right) R_{AA}^\alpha\left(\frac{p_\pi}{z},\phi\right) D_{\alpha\rightarrow \pi}\left(z,\frac{p_\pi}{z}\right)}
{\sum\limits_{\alpha=q,g}\;\int\limits_{z_{min}}^1 \frac{dz}{z} \; d\sigma_\alpha\left(\frac{p_\pi}{z}\right)  D_{\alpha\rightarrow \pi}\left(z,\frac{p_\pi}{z}\right)}\;\; ,
\label{RAApi}
\end{eqnarray}
\end{widetext}
with $z_{min}=\frac{2 p_\pi}{\sqrt{s}}$. The parton level 
$R_{AA}^\alpha$ is averaged over the initial jet production geometry in a given centrality class.
We use the KKP pion fragmentation functions \cite{Kniehl:2000hk}
that have been successfully tested on the $pp\rightarrow\pi^0$ spectra at RHIC and LHC.

Figure \ref{fig1} shows the partonic level $R_{AA}$ and $v_2$ as a function of $p_T$ 
for gluons (red dashed-dotted lines) and quarks (blue dashed lines) 
prior to fragmentation, as well as for pions (black solid lines) 
after fragmentation for a ($a=1/3,z=1,c=8/3)$ polytrope energy loss and 
Glauber initial conditions. The upper panel displays RHIC 
and the lower panel LHC energies where the coupling $\kappa$ was reduced 
to fit the central 0-5\% $R_{AA}^{\rm LHC}$ reference point at $ p_T=10$~GeV. 
The purpose of this figure is simply to demonstrate the remarkable similarity of quenched pion
and quenched quark jets at both RHIC and LHC in spite of 
the much higher number of initial gluon jets especially at LHC.

As can be seen from Fig.\ \ref{fig1}, the pion nuclear modification factor
is completely dominated by quark quenching and fragmentation at both RHIC
and LHC energies due to the fact that the gluon-energy loss is enhanced by a factor of 
$9/4$. The dominance of the quenched quark fragmentation reduces the sensitivity to 
the fragmentation function uncertainties because quark fragmentation functions
are the best constrained experimentally.

From the calculated $R_{AA}^\pi(p_\pi,\phi)=R_{AA}^\pi(p_\pi,N_{part},\phi)$
one obtains the centrality dependent $R_{AA}^\pi(N_{part})$ for a given 
$p_\pi$ by averaging over the azimuthal angles $\phi$ \cite{Betz:2011tu}. 
Then, the $v_2(N_{part})$ is computed via
\begin{eqnarray}
\hspace*{-2ex}
v_2^\pi(N_{part})
&=&\frac{\int d\phi \cos\left\{2\phi \right\} \,R_{AA}^\pi(N_{part},\phi)}
{\int d\phi \,R_{AA}^\pi(N_{part},\phi)} \,.
\end{eqnarray}

We fix the value of the jet-medium coupling $\kappa$ at RHIC energies in each case 
by fitting to the same one reference point with $R_{AA}(p_T=7.5\;{\rm GeV})\approx 0.2$ 
in central 0-5\% collisions.

\begin{figure}[t]
\hspace*{-1cm}
\vspace*{-0.3cm}
\includegraphics[scale=0.5]{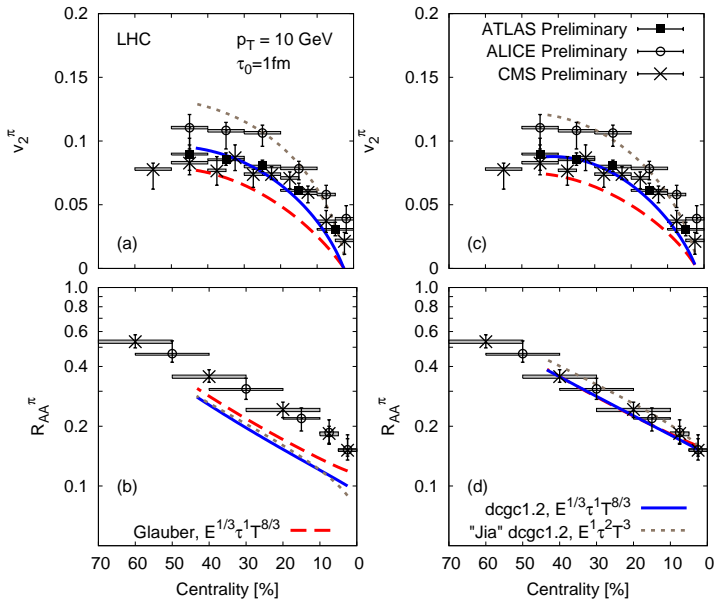}
\caption{(Color online) The $R^\pi_{AA}({\rm Centr.})$ and the 
$v^\pi_2({\rm Centr.})$ at LHC energies after fragmentation, considering
either the same coupling as for RHIC energies (left panel),
or a reduced coupling (right panel). Here we compare Glauber initial conditions
(red dashed lines) and dcgc1.2 initial conditions (blue solid lines) 
for an ($a=1/3,z=1,c=8/3$) energy-loss with a scenario for dcgc1.2 initial 
conditions considering just binary collisions (``Jia'' dcgc1.2) and ($a=1,z=2,c=3$) 
(grey dotted lines). Data are taken from Refs.\ 
\cite{CMSPas,Otwinowski:2011gq1,Abelev:2012di,Collaboration:2011hf,CMSRAA,CMSvn}.} 
\label{fig4}
\end{figure}

Glauber participant and binary collision initial condition geometries are computed 
using the Monte Carlo model introduced in Ref.\ \cite{Betz:2011tu}. To simulate 
CGC-like higher eccentricity initial conditions, we simply deform the  
Glauber initial geometry via the rescaling:
\begin{eqnarray}
x\rightarrow s_x x, \quad y \rightarrow s_y y\,,
\label{fkln1}
\end{eqnarray}
where the scaling factors are determined by fitting tabulated
Glauber and MC-KLN second moments as a function of $b$: 
\begin{eqnarray}
s_x = \sqrt{\frac{\langle x^2\rangle_{\rm CGC}}{\langle x^2\rangle_{\rm Gl}}}, 
\quad 
s_y = \sqrt{\frac{\langle y^2\rangle_{\rm CGC}}{\langle y^2\rangle_{\rm Gl}}}\,.
\end{eqnarray}
Here, $\langle\circ\rangle$ denotes the geometric average at a given $b$. 
Assuming that the ratios of eccentricities ($\epsilon=e_2$) and the
root-mean-square ($r^2=<x^2+y^2>$) for Glauber and CGC initial conditions
can be expressed via 
$\epsilon_{\rm CGC} = f\cdot\epsilon_{\rm Gl}$ and 
${\rm r}^2_{\rm CGC} = g\cdot {\rm r}^2_{\rm Gl}$, 
we found that deformations with $f=1.2\pm 0.1$ and $g=0.95\pm 0.05$ 
reproduce the numerical MC-KLN tables of Jia very well \cite{Jia2012}.

In the following, we will refer to the deformed CGC geometry corresponding to $(f=1.2,g=0.95)$
as ``dcgc1.2''. Please note that the Glauber geometry applied corresponds to 86\% participant 
plus 14\% binary fraction. Given $(f,g)$, the deformed CGC geometry has mean in-plane and 
out-of-(reaction)-plane moments of
\begin{eqnarray}
\langle x^2\rangle_{\rm CGC} &=& {g}\; {\rm r}^2_{\rm Gl} (1-f\;\epsilon_{\rm Gl})/2\nonumber\\
\langle y^2\rangle_{\rm CGC} &=& {g}\; {\rm r}^2_{\rm Gl} (1+f\;\epsilon_{\rm Gl})/2\;.
\end{eqnarray}

The energy-density field transforms under this elliptic deformation as
\begin{eqnarray}
\tilde{\epsilon}(x,y) &=& \epsilon\left(\frac{x}{s_x},\frac{y}{s_y}\right) \left(\frac{1}{s_x s_y} \right)^{4/3}\,.
\label{fkln2}
\end{eqnarray}
We checked that the temperature profiles and eccentricities 
of this azimuthally deformed Glauber model 
coincide with the ones of the fKLN model of Drescher et.\ al.\ 
\cite{adrian,dumitru}. 

In addition, we will also consider a QGP geometry that corresponds to deformed Glauber initial conditions
that are based on pure binary collisions which leads to a larger eccentricity. We will refer to that 
enhanced eccentricity geometry as ``Jia'' dcgc1.2 as this geometry, together with $a=1$, reproduces 
the PHENIX $v_2$-data reviewed below.

Since we showed in Ref.\ \cite{Betz:2011tu}
that event-by-event geometrical fluctuations lead to very similar
$R_{AA}$ and $v_2$ as event-averaged geometries, we consider only 
the event-averaged geometries in this paper.

\begin{figure*}
\begin{minipage}{4.4cm}
\hspace*{-2.2cm}
\includegraphics[scale=0.4]{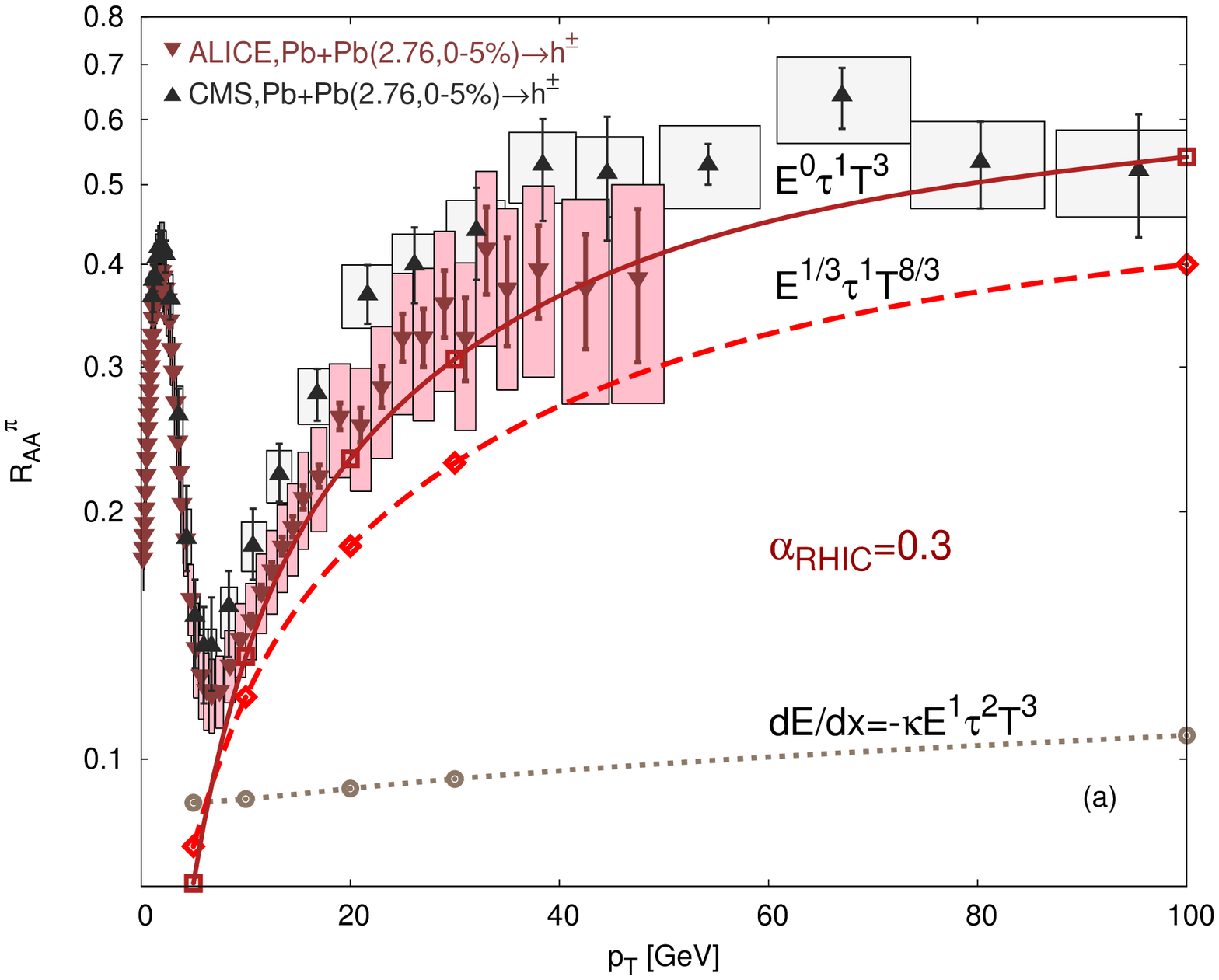}
\end{minipage}
\begin{minipage}{8.9cm}
\hspace*{1.8cm}
\includegraphics[scale=0.4]{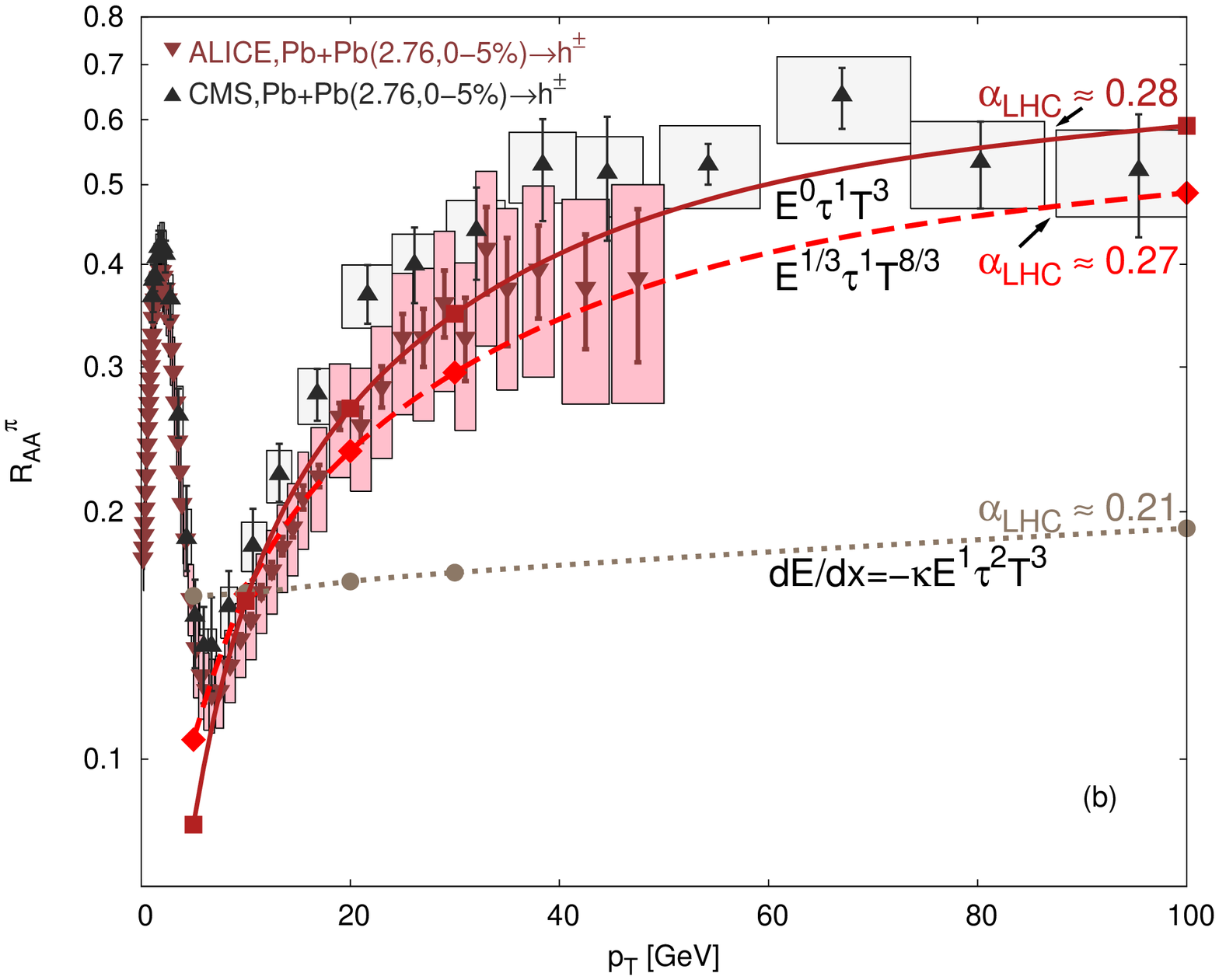}
\end{minipage}
\caption{(Color online) The nuclear modification factor of pions
at $0-5\%$ centrality as a function of $p_T$, shown for LHC energies,
assuming either the same coupling as at RHIC energies (left panel)
or a reduced coupling (right panel). The plot contrasts an energy-loss 
scenario for dcgc1.2 initial conditions considering just binary collisions 
(``Jia'' dcgc1.2, grey dotted lines) for $(a=1,z=2,c=3)$ with 
energy-loss scenarios for Glauber initial conditions and ($a=1/3,z=1,c=8/3$; 
red dashed lines) or ($a=0,z=1,c=3$; dark red solid lines), respectively.
The data are taken from Refs.\ \cite{CMSPas,Otwinowski:2011gq1,CMSRAA}.}
\label{fig5}
\end{figure*}

\section{Results and discussion}

Figure \ref{fig2} shows the centrality dependence of $R^\pi_{AA}(N_{part})$ 
(lower panels) and $v^\pi_2(N_{part})$ (upper panels) for pions with
$p_T=7.5$~GeV at RHIC energies considering a polytrope energy-loss scenario 
with $z=1$ (left panel) and $z=2$ (right panel) for three $(a,z,c=2-a+z)$ models. Red dashed
lines correspond to Glauber initial conditions for a pQCD-like scenario with 
$(a=1/3,z=1,c=8/3)$ as shown in the left panel, 
while a hybrid quadratic path-length dependence with $(a=1/3,z=2,c=11/3)$ is displayed in 
the right panels. Note that there is virtually no dependence on the $z$ exponent and 
for both scenarios, the results for the elliptic flow fall below the PHENIX data. 
The same polytropes for the deformed $(f=1.2,g=0.95)$ dcgc1.2 CGC-like geometry
are shown in solid blue. The enhanced eccentricity increases the predicted
$v_2(p_T=7.5\,{\rm GeV},N_{part})$ but the solid blue curves remain well below
the $v_2$-data. Therefore, neither geometries nor linear ($z=1$) or quadratic $(z=2)$ path
dependences account for the PHENIX data considering our default pQCD-like  $E^{1/3}$ polytrope case.

The long dashed-dotted black curves represents to the results of the ``survival probability'' 
model used in Refs. \cite{Jia,Adare:2010sp}. That model fits remarkably 
well both the centrality dependence of $R_{AA}$ and
$v_2$ simultaneously at the reference of $p_T=7.5$~GeV. 
This model, corresponds to the doubly special limit [Eq.\ (\ref{a1limlim})]
of the more general ``geometrical optics'' $(a=1, z=2, c=3)$ 
case given by Eq.\ (\ref{a1lim}). However, since the spectral index is not constant as
a function of $p_T$ at RHIC, this double limit is not consistent.

Therefore, we also show the dotted grey lines in Fig.\ \ref{fig2} using the full spectral shape
with  Eq.\ (\ref{RAApi}) and corresponding to $dE/dx\propto E^1$. 
Applying the high eccentricity ``Jia'' dcgc1.2 geometry of Refs.\ \cite{Jia,Adare:2010sp}, 
we find that Eq.\ (\ref{RAApi}) results in $v_2(p_T=7.5\,{\rm GeV},N_{part})$ close to Jia's curve and the data.

The results for $a=1/3$ are comparable to those shown in Ref.\ \cite{Betz:2011tu} 
where, however, the full convolution over the parton spectra 
and pion fragmentation functions was not performed. 

We note that the fKLN model used in Ref.\ \cite{Betz:2011tu}
appeared to reproduce the data for $z=2$, but we discovered 
that the numerics of the fKLN code used was not stable to small
variations of parameter settings. This is the reason we abandoned the
fKLN calculations in this work and chose instead to compare Glauber
to deformed Glauber dcgc1.2 geometries.
Our conclusion with new deformed geometry is that $a=1/3$
polytropes underestimate $v_2$ at RHIC independent of path-length dependence.

If the simple geometric optics model with $(a=1)$ is the correct explanation of $v_2$-data,
then the $R_{AA}(p_T)$ should be approximately independent of $p_T$ as can be seen
from Eq.\ (\ref{RAApi}). In Fig.\ \ref{fig3} this prediction is compared to 
RHIC data. We note that the curve for $(a=1,z=2,c=3)$ deceases slightly due to using the
full spectral shape as well as the convolution over the pion fragmentation
functions. Current RHIC data are indeed consistent with an approximately $p_T$ independence
in the $5<p_T<10$~GeV window, but the large systematic errors beyond $10$~GeV prevent a stringent
test of this flatness. As seen from the figure, $a=0,1/3$ energy exponents are 
consistent within the error bars. Future higher statistics measurements at RHIC in the range $5<p_T<30$~GeV 
are obviously needed to differentiate between the energy-loss models.
 
Having constrained the coupling of different polytrope models at RHIC,
we turn to the predictions of these models for LHC conditions
in Figs.\ \ref{fig4} and \ref{fig5}. Here, the LHC data are taken from Refs.\ 
\cite{CMSPas,Abelev:2012di,CMSvn}.
In Eq.\ (\ref{RAApi}) we use the parton spectra predicted by pQCD as well as 
the doubling of the density between RHIC and LHC. We scale the temperature 
field used in Eq.\ (\ref{FinMom}) by a factor $(2.2)^{1/3}$ relative to RHIC. 

Fig.\ \ref{fig4}(b) is one of the main results showing that all three RHIC 
constrained polytrope models shown in Fig.\ \ref{fig3} overpredict the
jet suppression at LHC for most central collisions at the LHC-reference point of 
$p_T=10$~GeV. Thus, we find in Fig.\ \ref{fig4}(b) that our extrapolation from RHIC to 
LHC conditions assuming the same jet-medium coupling as the density increases by a factor 
of $\sim 2.2$ from RHIC to LHC underpredicts the observed $R_{PbPb}(p_T=10\,{\rm GeV},0-5\%)$ 
at all centralities and that this discrepancy is robust to substantial variations of the 
$(a,z,c)$ exponents of the energy-loss model assumed. This results extends the robustness
of the evidence for a reduced jet-medium coupling at LHC found in Refs.\ 
\cite{Horowitz:2011gd,Zakharov:2011ws,Buzzatti:2011vt} to a much
broader class of models.

The main result found in Fig.\ \ref{fig2} was that the measured $R_{AA}^\pi(N_{\rm part})$ 
and $v_2^\pi(N_{\rm part})$ at RHIC at $p_T=7.5$~GeV appears to be best described with 
a polytrope model for $(a=1,z=2,c=3)$. The predicted $p_T$-flatness of the nuclear
modification factor shown in Fig.\ \ref{fig3} is also consistent with the measured data.
However, we by comparing with LHC data in Fig.\ \ref{fig5} we learn that 
the $dE/dx\propto E^1$ dynamics is completely falsified by the rising $R_{AA}(p_T)$ 
at RHIC energies which is independent of the magnitude of the jet-medium coupling.

\begin{figure*}
\begin{minipage}{4.4cm}
\hspace*{-1.2cm}
\includegraphics[scale=0.4]{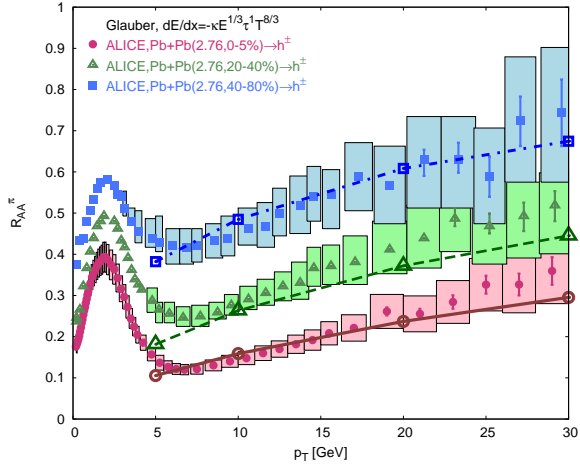}
\end{minipage}
\caption{(Color online) The nuclear modification factor of pions
as a function of $p_T$ for $0-5\%$ (solid magenta line), 
$20-40\%$ (dashed green line), and $40-80\%$ centralities (dashed-dotted blue line) 
at LHC energies, compared to an energy-loss scenario of
$(a=1/3,z=1,c=8/3)$. Results are computed assuming  Glauber 
geometry but $R_{AA}$ curves with dcgc1.2 (not shown) 
were found to be essentially the same. The centrality classes are computed with
the same reduced coupling determined from the most central $p_T=10$~GeV reference point. The data are taken from Ref.\cite{Appelshauser:2011ds}.}
\label{fig6}
\end{figure*}

Fig. 5b, in which the jet-medium couplings of the three different polytropes are reduced
to fit the $10$~GeV reference point at most central collisions, clearly shows that,
within the current experimental uncertainties, the rapid rise of the $R_{AA}(p_T)$ 
at the LHC rules out any energy-loss model with an exponent $a>1/3$. 
In fact, the data appear to slightly favour the $a=0$ case.

As a further consistency test we compare the centrality dependence of the
$R_{AA}^{Pb+Pb}(5<p_T<30\,{\rm GeV})$ for the polytrope model with $(a=1/3,z=1,c=8/3)$ 
in Fig.\ \ref{fig6} \cite{Otwinowski:2011gq1} to ALICE data. The observed agreement 
between data and model calculations is in fact to robust to changes between 
Glauber and deformed dcgc1.2 geometries for initial conditions.

Since our results support the conclusion that the QGP created at LHC appears more 
transparent than expected based on fixed-coupling extrapolations from RHIC
\cite{Horowitz:2011gd,Buzzatti:2011vt}, it is important to try to quantify
the magnitude of the reduction needed to obtain the agreement shown in  Fig.\ \ref{fig4}(d).  
The ratio of the coupling needed to reproduce the LHC reference point to the one that fits
the RHIC data provides a useful measure of the degree of weakening of the effective 
jet-medium coupling as the QGP density doubles from RHIC to LHC.

In pQCD radiative energy loss, $\kappa\propto \alpha^3$ 
and the strong coupling thus scales from RHIC to LHC as
\begin{eqnarray}
\alpha_{\rm LHC}&=& (\kappa_{\rm LHC}/\kappa_{\rm RHIC})^{1/3} \; \alpha_{\rm RHIC}\,,
\end{eqnarray}
where $\alpha_{\rm RHIC}\sim 0.3$. Inserting the values used in Fig.\ 
\ref{fig4}, which are summarized in Table \ref{tablekappa1} for the $z=1$ energy loss
shown as well as a $z=2$ energy loss not shown here in detail, 
we find that $\alpha_{\rm LHC}\sim 0.24 - 0.27$ for an energy-loss scenario
of $a=1/3$, indicating a plausible moderate reduction of the pQCD coupling 
due to slow running (creeping) above the deconfinement temperature (see also Ref.\ 
\cite{Horowitz:2011gd,Zakharov:2011ws,Buzzatti:2011vt}), while for the "Jia"
optical limit $\alpha_{\rm LHC}\sim 0.21$ and for $(a=0,z=1,c=3)$ $\alpha_{\rm LHC}\sim 0.28$. 
Remarkably, by comparing the values for $a=1/3$ in Table \ref{tablekappa1} with Table 
\ref{tablekappa2}, we find that the ratio of LHC to RHIC effective couplings
is insensitive to the assumed $\tau_0$ in the range 0-1 fm/c.

\begin{table}[b]
\begin{tabular}{|c|c|c|c|c|c|}
\hline
\multicolumn{6}{|c|}{Effective Coupling $\kappa$ assuming $\tau_0=1.0$ fm/c}\\
\hline
$\sqrt{s}$ & Glauber & dcgc1.2 & Glauber & Glauber & "Jia"\\
  &  a=1/3 & a=1/3 & a=1/3 & a=0 & a=1\\
  &  z=1 & z=1 & z=2 & z=1 & z=2\\
\hline
0.20  & 0.93 &  1.09  & 0.55 & 3.30 & 0.057 \\
\hline
2.76  & 0.66 & 0.66  & 0.33 & 2.72 & 0.017\\
\hline
\hline
LHC/RHIC & 0.71 & 0.61 & 0.60 & 0.82 & 0.33\\
\hline
\end{tabular}
\caption{The effective coupling $\kappa$ at RHIC and LHC energies 
for Glauber and dcgc1.2 initial conditions and $\tau_0=1.0$ fm/c.
The last row displays the ratio $\kappa_{LHC}/\kappa_{RHIC}$.}
\label{tablekappa1}
\end{table}

On the other hand, in the falling-string scenario 
\cite{ches1,ches2,Arnold:2011qi,Ficnar:2011yj,Ficnarnew}, the effective 
jet-medium coupling $\kappa\propto \sqrt{\lambda}$ is related to the 
square root of the t'Hooft coupling $\lambda=g_{YM}^2 N_c$. Gravity-dual 
descriptions require that $\lambda\gg 1$. For heavy-quark quenching, it was 
found in Ref.\ \cite{coester} that large $\lambda_{\rm RHIC} \sim 20$ 
provides a reasonably good fit to the RHIC data as well as to the bulk 
$v_2$ elliptic flow. 

In the falling string scenario, $\lambda_{\rm LHC}$ and 
$\lambda_{\rm RHIC}$ are then related via
\begin{eqnarray}
\lambda_{\rm LHC}&=& (\kappa_{\rm LHC}/\kappa_{\rm RHIC})^2 \; \lambda_{\rm RHIC}\,.
\end{eqnarray}
From the holographic point of view, Tables \ref{tablekappa1} and 
\ref{tablekappa2} imply that $\lambda_{\rm LHC}$ must be reduced 
by a rather large factor of $\sim 2-4$ relative to RHIC 
for the $a=1/3$ scenario and an even larger factor of $\sim 10$ 
for the optical "Jia" limit. 
This result implies a rather strong breaking 
of conformal symmetry over a narrow temperature interval. It is not yet clear
if current non-conformal holographic models are consistent with such a 
strong variation (see, for example, Refs.\ \cite{Ficnar:2011yj,Mia:2011iv}).

\begin{figure*}
\hspace*{-0.7cm}
\includegraphics[scale=0.45]{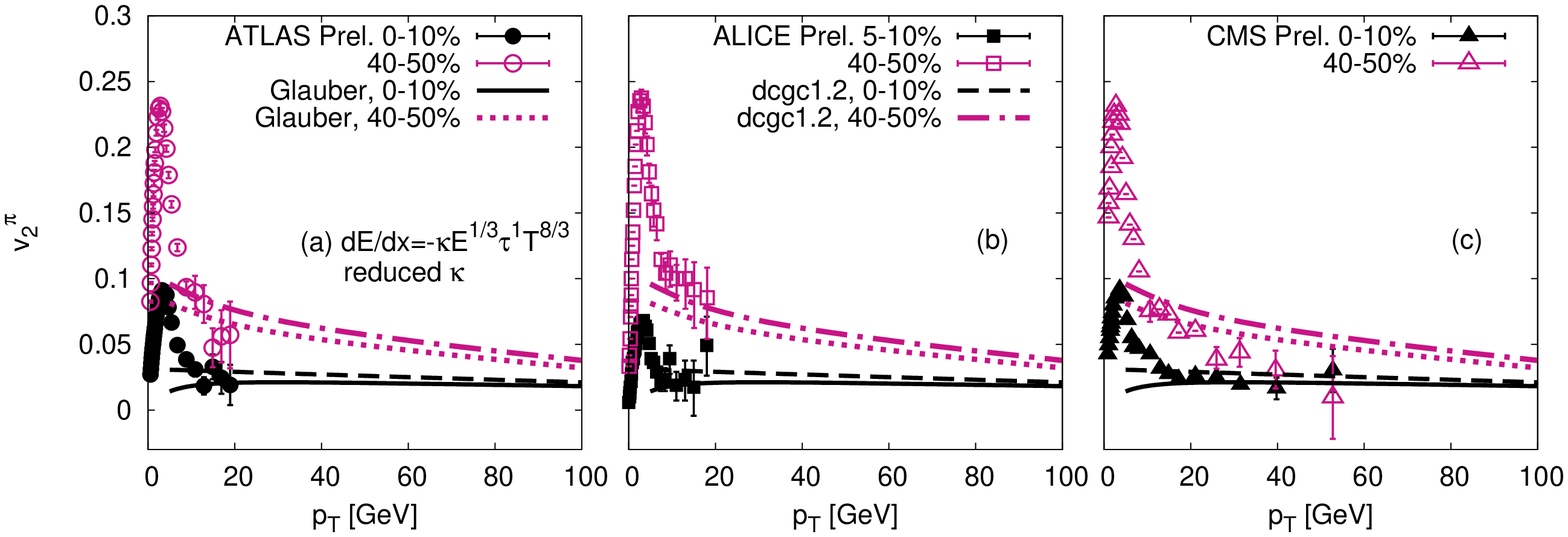}
\caption{(Color online) $v^\pi_2$ as a function of $p_T$, shown for 
different centralities at LHC energies for an energy loss with $(a=1/3,z=1,c=8/3)$, 
compared to ATLAS (left panel), ALICE (middle panel), and CMS data
(right panel) for central (full symbols) and peripheral collisions (open symbols). 
Glauber initial conditions are displayed by the solid black and dotted magenta lines 
while the dcgc1.2 initial conditions are represented by the dashed black and dashed-dotted magenta
lines. Here, the initialization time is chosen to be $\tau_0=1$~fm and 
a reduced coupling constant is assumed as compared to RHIC energies.
The data are taken from Refs.\ \cite{Abelev:2012di,Collaboration:2011hf,CMSvn}.}
\label{fig7}
\end{figure*}

\begin{table}[t]
\begin{tabular}{|c|c|c|c|}
\hline
\multicolumn{4}{|c|}{Effective Coupling $\kappa$ assuming $\tau_0=0.01$ fm/c}\\
\hline
$\sqrt{s}$ & Glauber & dcgc1.2 & Glauber \\
  &  z=1 & z=1 & z=2\\
\hline
0.20  & 0.60 &  0.58  & 0.44\\
\hline
2.76  & 0.45 & 0.43  & 0.26\\
\hline
\hline
LHC/RHIC & 0.75 & 0.74 & 0.59\\\hline
\end{tabular}
\caption{Same as table I but for fits assuming $\tau_0=0.01$ fm/c.}
\label{tablekappa2}
\end{table}

Figs.\ \ref{fig2} and \ref{fig4} already showed that there is a
remarkably insensitivity of the elliptic flow to the initial conditions
and the strength of the coupling constant [compare e.g.\ Fig.\ \ref{fig4}(a)
with Fig.\ \ref{fig4}(c)]. Fig.\ \ref{fig7} demonstrates that
elliptic flow measurements at LHC, which recently are in remarkable
agreement between ATLAS, ALICE, and CMS \cite{Otwinowski:2011gq1,Abelev:2012di,Collaboration:2011hf,CMSRAA,CMSvn}
will not allow for a disentangling of Glauber vs.\ dcgc1.2 initial conditions. 
Preliminary data from CMS up to $p_T \sim 50$~GeV \cite{CMSvn}
(right panel of Fig.\ \ref{fig7}) show that the 
the elliptic flow for central collisions indeed stays flat 
for high momenta while $v^\pi_2$ for more peripheral collisions 
decreases slowly approaching the value for central collisions.

\section{Summary}

Using a generic, polytrope power-law model of energy, path-length,
and monotonic density dependence characterized by the three exponents
$(a,z,c)$ that can interpolate between a wide class of weakly and 
strongly-coupled jet-medium interactions, we investigated 
those interactions as well as the jet-medium coupling at both RHIC and 
LHC energies for Glauber and CGC-like, deformed Glauber (dcgc1.2) initial conditions.

We found that at RHIC energies the measured data for the nuclear modification
factor and the elliptic flow as a function of centrality are best described 
by the scenario with $(a=1,z=2,c=3)$ that also describes the flatness of the
$R_{AA}^{\rm RHIC}(p_T)$.

However, this energy-loss model is completely falsified at LHC energies
because it does not describe the rising of the $R_{AA}^{\rm LHC}(p_T)$.
In fact, the rapid rise of the nuclear modification factor rules out any
energy-loss model with an energy exponent $a>1/3$ and slightly favours 
the $a=0$ case.

Moreover, we find that an extrapolation from RHIC to LHC energies
leads to an underprediction of the nuclear modification factor 
\cite{Horowitz:2011gd,Zakharov:2011ws, Buzzatti:2011vt}, independently
of the $(a,z,c)$ exponents assumed, emphasizing the robust
evidence for a reduced jet-medium coupling at the LHC.
We showed that in terms of pQCD a moderate reduction of the effective jet-medium coupling 
with $\alpha_{LHC}\sim 0.24-0.28$ relative to RHIC
($\alpha_{RHIC}=0.3$) allows to describe both the nuclear modification factor
and the elliptic flow as a function of centrality and as a function of $p_T$. Thus,
the LHC data are compatible with $0\leq a\leq 1/3$ pQCD-like energy-loss models where the
jet-medium coupling is reduced by approximately 10\% between RHIC and LHC.

In terms of strongly-coupled holographic models, however, our LHC fit 
requires a much larger reduction of the effective t'Hooft coupling 
from $\lambda_{\rm RHIC} \sim 20$ by a factor of $2-4$ to 
$\lambda_{\rm LHC}\sim 5-10$.  This suggests that stronger non-conformal effects 
must be considered for a holographic phenomenology of the LHC.

We found that both Glauber and CGC-like, deformed Glauber (dcgc1.2) initial conditions describe
the centrality dependence of the nuclear modification factor 
$R^\pi_{AA}(p_T)$ and the elliptic flow $v^\pi_2$ well for RHIC and 
LHC energies which does not allow us to disentangle the initial 
conditions using those two observables.

As demonstrated in Fig.\ \ref{fig1}, the nuclear modification factor 
as well as the elliptic flow at both RHIC and LHC energies for 
$p_T>5$ GeV are dominated by quark jet-quenching and fragmentation 
because gluon jets are strongly quenched and fragmentation leads
to pions with a smaller fractional momentum.
This underlines the conclusion of Ref.\ \cite{Buzzatti:2011vt}
that single-hadron jet-flavour tomography observables are mainly sensitive
to quark rather than gluon jet-medium interactions. 
The reduced jet-medium coupling quantified in this work therefore primarily
refers to the apparent weakening of quark-jet interactions in a QGP 
when the density approximately doubles from RHIC to LHC. It remains 
a challenge to identify jet observables more sensitive to gluon 
jet-quenching to test color Casimir scaling currently assumed in both 
weakly-coupled pQCD tomography and strongly-coupled string holography. 
Di-hadron and jet-shape observables could help to probe gluon versus quark 
jet-medium interactions in the future.

\section*{Acknowledgements}
M.G. and B.B.\ acknowledge support from DOE under
Grant No.\ DE-FG02-93ER40764. B.B.\ is supported by the Alexander von Humboldt 
foundation via the Feodor Lynen program. The authors thank G.\ Torrieri, J.\ Jia,
A.\ Buzzatti, A.\ Ficnar, W.\ Horowitz, J. Liao, M.\ Mia, J.\ Noronha, 
J.\ Harris, G.\ Roland, B.\ Cole, and W.\ Zajc
for extensive and fruitful discussions.
\\

\end{document}